# Network Embedding as Matrix Factorization: Unifying DeepWalk, LINE, PTE, and node2vec

Jiezhong Qiu[†*], Yuxiao Dong[‡], Hao Ma[‡], Jian Li[♯], Kuansan Wang[‡], and Jie Tang[†]

[†]Department of Computer Science and Technology, Tsinghua University
[‡]Microsoft Research, Redmond
[♯]Institute for Interdisciplinary Information Sciences, Tsinghua University
qiujz16@mails.tsinghua.edu.cn,{yuxdong,haoma,kuansanw}@microsoft.com,{lijian83,jietang}@tsinghua.edu.cn

## ABSTRACT

Since the invention of word2vec [28, 29], the skip-gram model has significantly advanced the research of network embedding, such as the recent emergence of the DeepWalk, LINE, PTE, and node2vec approaches. In this work, we show that all of the aforementioned models with negative sampling can be unified into the matrix factorization framework with closed forms. Our analysis and proofs reveal that: (1) DeepWalk [31] empirically produces a low-rank transformation of a network's normalized Laplacian matrix; (2) LINE [37], in theory, is a special case of DeepWalk when the size of vertices' context is set to one; (3) As an extension of LINE, PTE [36] can be viewed as the joint factorization of multiple networks' Laplacians; (4) node2vec [16] is factorizing a matrix related to the stationary distribution and transition probability tensor of a 2nd-order random walk. We further provide the theoretical connections between skip-gram based network embedding algorithms and the theory of graph Laplacian. Finally, we present the NetMF method[1] as well as its approximation algorithm for computing network embedding. Our method offers significant improvements over DeepWalk and LINE for conventional network mining tasks. This work lays the theoretical foundation for skip-gram based network embedding methods, leading to a better understanding of latent network representation learning.

[*]This work was partially done when Jiezhong was an intern at Microsoft Research.
[1]Code available at github.com/xptree/NetMF.

**Table 1: The matrices that are implicitly approximated and factorized by DeepWalk, LINE, PTE, and node2vec.**

| Algorithm | Matrix |
|---|---|
| DeepWalk | $\log\left(\text{vol}(G)\left(\frac{1}{T}\sum_{r=1}^{T}(D^{-1}A)^r\right)D^{-1}\right) - \log b$ |
| LINE | $\log\left(\text{vol}(G)D^{-1}AD^{-1}\right) - \log b$ |
| PTE | $\log\left(\begin{bmatrix}\alpha\,\text{vol}(G_{\text{ww}})(D_{\text{row}}^{\text{ww}})^{-1}A_{\text{ww}}(D_{\text{col}}^{\text{ww}})^{-1}\\\beta\,\text{vol}(G_{\text{dw}})(D_{\text{row}}^{\text{dw}})^{-1}A_{\text{dw}}(D_{\text{col}}^{\text{dw}})^{-1}\\\gamma\,\text{vol}(G_{\text{lw}})(D_{\text{row}}^{\text{lw}})^{-1}A_{\text{lw}}(D_{\text{col}}^{\text{lw}})^{-1}\end{bmatrix}\right) - \log b$ |
| node2vec | $\log\left(\frac{\frac{1}{2T}\sum_{r=1}^{T}\left(\sum_u X_{w,u}\underline{P}_{c,w,u}^r + \sum_u X_{c,u}\underline{P}_{w,c,u}^r\right)}{(\sum_u X_{w,u})(\sum_u X_{c,u})}\right) - \log b$ |

Notations in DeepWalk and LINE are introduced below. See detailed notations for PTE and node2vec in Section 2.
$A$: $A \in \mathbb{R}_+^{|V|\times|V|}$ is $G$'s adjacency matrix with $A_{i,j}$ as the edge weight between vertices $i$ and $j$;
$D_{\text{col}}$: $D_{\text{col}} = \text{diag}(A^\top e)$ is the diagonal matrix with column sum of $A$;
$D_{\text{row}}$: $D_{\text{row}} = \text{diag}(Ae)$ is the diagonal matrix with row sum of $A$;
$D$: For undirected graphs ($A^\top = A$), $D_{\text{col}} = D_{\text{row}}$. For brevity, $D$ represents both $D_{\text{col}}$ & $D_{\text{row}}$.
$D = \text{diag}(d_1, \cdots, d_{|V|})$, where $d_i$ represents generalized degree of vertex $i$;
$\text{vol}(G)$: $\text{vol}(G) = \sum_i \sum_j A_{i,j} = \sum_i d_i$ is the volume of a weighted graph $G$;
$T$ & $b$: The context window size and the number of negative sampling in skip-gram, respectively.

centrality, triangle count, and modularity, require extensive domain knowledge and expensive computation to handcraft. In light of these issues, as well as the opportunities offered by the recent emergence of representation learning [2], learning latent representations for networks, a.k.a., network embedding, has been extensively studied in order to automatically discover and map a network's structural properties into a latent space.

Formally, the problem of network embedding is often formalized as follows: Given an undirected and weighted graph $G = (V, E, A)$ with $V$ as the node set, $E$ as the edge set and $A$ as the adjacency matrix, the goal is to learn a function $V \to \mathbb{R}^d$ that maps each vertex to a $d$-dimensional ($d \ll |V|$) vector that captures its structural properties. The output representations can be used as the input of mining and learning algorithms for a variety of network science tasks, such as label classification and community detection.

The attempt to address this problem can date back to spectral graph theory [11] and social dimension learning [38]. Its very recent advances have been largely influenced by the skip-gram model originally proposed for word embedding [28, 29], whose input is a text corpus composed of sentences in natural language and output is the latent vector representation for each word in the corpus. Notably, inspired by this setting, DeepWalk [31] pioneers network embedding by considering the vertex paths traversed by random walks over networks as the sentences and leveraging skip-gram for learning latent vertex representations. With the advent of

## 1 INTRODUCTION

The conventional paradigm of mining and learning with networks usually starts from the explicit exploration of their structural properties [13, 32]. But many of such properties, such as betweenness

DeepWalk, many network embedding models have been developed, such as LINE [37], PTE [36], and node2vec [16].

The above models have thus far been demonstrated quite effective empirically. However, the theoretical mechanism behind them is much less well-understood. We note that the skip-gram model with negative sampling for word embedding has been shown to be an implicit factorization of a certain word-context matrix [24], and there is recent effort to theoretically explaining the word embedding models from geometric perspectives [1, 18]. But it is unclear what is the relation between the word-context matrix and the network structure. Moreover, there was an early attempt to theoretically analyze DeepWalk's behavior [47]. However, their main theoretical results are not fully consistent with the setting of the original DeepWalk paper. In addition, despite the superficial similarity among DeepWalk, LINE, PTE, and node2vec, there is a lack of deeper understanding of their underlying connections.

**Contributions** In this work, we provide theoretical results concerning several skip-gram powered network embedding methods. More concretely, we first show that the models we mentioned—DeepWalk, LINE, PTE, and node2vec—are in theory performing implicit matrix factorizations. We derive the closed form of the matrix for each model (see Table 1 for a summary). For example, DeepWalk (random walks on a graph + skip-gram) is in essence factorizing a random matrix that converges in probability to our closed-form matrix as the length of random walks goes to infinity.

Second, observed from their matrices' closed forms, we find that, interestingly, LINE can be seen as a special case of DeepWalk, when the window size $T$ of contexts is set to 1 in skip-gram. Furthermore, we demonstrate that PTE, as an extension of LINE, is actually an implicit factorization of the joint matrix of multiple networks.

Third, we discover a theoretical connection between DeepWalk's implicit matrix and graph Laplacians. Based on the connection, we propose a new algorithm *NetMF* to approximate the closed form of DeepWalk's implicit matrix. By explicitly factorizing this matrix using SVD, our extensive experiments in four networks (used in the DeepWalk and node2vec approaches) demonstrate NetMF's outstanding performance (relative improvements by up to 50%) over DeepWalk and LINE.

## 2 THEORETICAL ANALYSIS AND PROOFS

In this section, we present the detailed theoretical analysis and proofs for four popular network embedding approaches: LINE, PTE, DeepWalk, and node2vec.

### 2.1 LINE and PTE

**LINE [37]** Given an undirected and weighted network $G = (V, E, A)$, LINE with the second order proximity (aka LINE (2nd)) aims to learn two representation matrices $X, Y \in \mathbb{R}^{|V| \times d}$, whose rows are denoted by $x_i$ and $y_i$, $i = 1, \cdots, |V|$, respectively. The objective of LINE (2nd) is to maximize

$$\ell = \sum_{i=1}^{|V|} \sum_{j=1}^{|V|} A_{i,j} \left( \log g\left(x_i^\top y_j\right) + b \mathbb{E}_{j' \sim P_N} \left[ \log g\left(-x_i^\top y_{j'}\right) \right] \right),$$

where $g$ is the sigmoid function; $b$ is the parameter for negative sampling; $P_N$ is known as the noise distribution that generates negative samples. In LINE, the authors empirically set $P_N(j) \propto d_j^{3/4}$, where $d_j = \sum_{k=1}^{|V|} A_{j,k}$ is the generalized degree of vertex $j$. In our analysis, we assume $P_N(j) \propto d_j$ because the normalization factor has a closed form solution in graph theory, i.e., $P_N(j) = d_j/\text{vol}(G)$, where $\text{vol}(G) = \sum_{i=1}^{|V|} \sum_{j=1}^{|V|} A_{i,j}$.[2] Then we rewrite the objective as

$$\ell = \sum_{i=1}^{|V|} \sum_{j=1}^{|V|} A_{i,j} \log g\left(x_i^\top y_j\right) + b \sum_{i=1}^{|V|} d_i \mathbb{E}_{j' \sim P_N} \left[ \log g\left(-x_i^\top y_{j'}\right) \right], \quad (1)$$

and express the expectation term $\mathbb{E}_{j' \sim P_N}\left[\log g\left(-x_i^\top y_{j'}\right)\right]$ as

$$\frac{d_j}{\text{vol}(G)} \log g\left(-x_i^\top y_j\right) + \sum_{j' \neq j} \frac{d_{j'}}{\text{vol}(G)} \log g\left(-x_i^\top y_{j'}\right). \quad (2)$$

By combining Eq. 1 and Eq. 2, and considering the local objective function for a specific pair of vertices $(i, j)$, we have

$$\ell(i, j) = A_{i,j} \log g\left(x_i^\top y_j\right) + b \frac{d_i d_j}{\text{vol}(G)} \log g\left(-x_i^\top y_j\right).$$

Let us define $z_{i,j} = x_i^\top y_j$. Following Levy and Goldberg [24], where the authors suggested that for a sufficient large embedding dimension, each individual $z_{i,j}$ can assume a value independence, we can take the derivative w.r.t. $z_{i,j}$ and get:

$$\frac{\partial \ell}{\partial z_{i,j}} = \frac{\partial \ell(i,j)}{\partial z_{i,j}} = A_{i,j} g(-z_{i,j}) - b \frac{d_i d_j}{\text{vol}(G)} g(z_{i,j}).$$

Setting the derivative to be zero reveals

$$e^{2z_{i,j}} - \left(\frac{\text{vol}(G) A_{i,j}}{b d_i d_j} - 1\right) e^{z_{i,j}} - \frac{\text{vol}(G) A_{i,j}}{b d_i d_j} = 0.$$

The above quadratic equation has two solutions (1) $e^{z_{i,j}} = -1$, which is invalid; and (2) $e^{z_{i,j}} = \frac{\text{vol}(G) A_{i,j}}{b d_i d_j}$, i.e.,

$$x_i^\top y_j = z_{i,j} = \log\left(\frac{\text{vol}(G) A_{i,j}}{b d_i d_j}\right). \quad (3)$$

Writing Eq. 3 in matrix form, LINE (2nd) is factoring the matrix

$$\log\left(\text{vol}(G) D^{-1} A D^{-1}\right) - \log b = X Y^\top, \quad (4)$$

where $\log(\cdot)$ denotes the element-wise matrix logarithm, and $D = \text{diag}(d_1, \cdots, d_{|V|})$.

**PTE [36]** PTE is an extension of LINE (2nd) in heterogeneous text networks. To examine it, we first adapt our analysis of LINE (2nd) to bipartite networks. Consider a bipartite network $G = (V_1 \cup V_2, E, A)$ where $V_1, V_2$ are two disjoint sets of vertices, $E \subseteq V_1 \times V_2$ is the edge set, and $A \in \mathbb{R}_+^{|V_1| \times |V_2|}$ is the bipartite adjacency matrix. The volume of $G$ is defined to be $\text{vol}(G) = \sum_{i=1}^{|V_1|} \sum_{j=1}^{|V_2|} A_{i,j}$. The goal is to learn a representation $x_i$ for each vertex $v_i \in V_1$ and a representation $y_j$ for each vertex $v_j \in V_2$. The objective function is

$$\ell = \sum_{i=1}^{|V_1|} \sum_{j=1}^{|V_2|} A_{i,j} \left(\log g\left(x_i^\top y_j\right) + b \mathbb{E}_{j' \sim P_N}\left[\log g\left(-x_i^\top y_{j'}\right)\right]\right).$$

Applying the same analysis procedure of LINE, we can see that maximizing $\ell$ is actually factorizing

$$\log\left(\text{vol}(G) D_{\text{row}}^{-1} A D_{\text{col}}^{-1}\right) - \log b = X Y^\top$$

where we denote $D_{\text{row}} = \text{diag}(Ae)$ and $D_{\text{col}} = \text{diag}(A^\top e)$.

Given the above discussion, let us consider the heterogeneous text network used in PTE, which is composed of three sub-networks

---
[2] A similar result could be achieved if we use $P_N(j) \propto d_j^{3/4}$.

— the word-word network $G_{ww}$, the document-word network $G_{dw}$, and the label-word network $G_{lw}$, where $G_{dw}$ and $G_{lw}$ are bipartite. Take the document-word network $G_{dw}$ as an example, we use $A_{dw} \in \mathbb{R}^{\#doc \times \#word}$ to denote its adjacency matrix, and use $D_{row}^{dw}$ and $D_{col}^{dw}$ to denote its diagonal matrices with row and column sum, respectively. By leveraging the analysis of LINE and the above notations, we find that PTE is factorizing

$$\log \left( \begin{bmatrix} \alpha \operatorname{vol}(G_{ww})(D_{row}^{ww})^{-1} A_{ww} (D_{col}^{ww})^{-1} \\ \beta \operatorname{vol}(G_{dw})(D_{row}^{dw})^{-1} A_{dw} (D_{col}^{dw})^{-1} \\ \gamma \operatorname{vol}(G_{lw})(D_{row}^{lw})^{-1} A_{lw} (D_{col}^{lw})^{-1} \end{bmatrix} \right) - \log b, \qquad (5)$$

where the factorized matrix is of shape (#word + #doc + #label) × #word, $b$ is the parameter for negative sampling, and $\{\alpha, \beta, \gamma\}$ are non-negative hyper-parameters to balance the weights of the three sub-networks. In PTE, $\{\alpha, \beta, \gamma\}$ satisfy $\alpha \operatorname{vol}(G_{ww}) = \beta \operatorname{vol}(G_{dw}) = \gamma \operatorname{vol}(G_{lw})$. This is because the authors perform edge sampling during training wherein edges are sampled from each of three sub-networks alternatively (see Section 4.2 in [36]).

## 2.2 DeepWalk

In this section, we analyze DeepWalk [31] and illustrate the essence of DeepWalk is actually performing an implicit matrix factorization (See the matrix form solution in Thm. 2.3).

DeepWalk first generates a "corpus" $\mathcal{D}$ by random walks on graphs [26]. To be formal, the corpus $\mathcal{D}$ is a multiset that counts the multiplicity of vertex-context pairs. DeepWalk then trains a skip-gram model on the multiset $\mathcal{D}$. In this work, we focus on skip-gram with negative sampling (SGNS). For clarity, we summarize the DeepWalk method in Algorithm 1. The outer loop (Line 1-7) specifies the total number of times, $N$, for which we should run random walks. For each random walk of length $L$, the first vertex is sampled from a prior distribution $P(w)$. The inner loop (Line 4-7) specifies the construction of the multiset $\mathcal{D}$. Once we have $\mathcal{D}$, we run an SGNS to attain the network embedding (Line 8). Next, we introduce some necessary background about the SGNS technique, followed by our analysis of the DeepWalk method.

**Preliminary on Skip-gram with Negative Sampling (SGNS)**
The skip-gram model assumes a corpus of words $w$ and their context $c$. More concretely, the words come from a textual corpus of words $w_1, \cdots w_L$ and the contexts for word $w_i$ are words surrounding it in a $T$-sized window $w_{i-T}, \cdots, w_{i-1}, w_{i+1}, \cdots, w_{i+T}$. Following the work by Levy and Goldberg [24], SGNS is implicitly factorizing

$$\log \left( \frac{\#(w,c) |\mathcal{D}|}{\#(w) \cdot \#(c)} \right) - \log b, \qquad (6)$$

where $\#(w,c)$, $\#(w)$ and $\#(c)$ denote the number of times word-context pair $(w,c)$, word $w$ and context $c$ appear in the corpus, respectively; $b$ is the number of negative samples.

**Proofs and Analysis** Our analysis of DeepWalk is based on the following key assumptions. Firstly, assume the used graph is undirected, connected, and non-bipartite, making $P(w) = d_w/\operatorname{vol}(G)$ a unique stationary distribution of the random walks. Secondly, suppose the first vertex of a random walk is sampled from the stationary distribution $P(w) = d_w/\operatorname{vol}(G)$.

To characterize DeepWalk, we want to reinterpret Eq. 6 by using graph theory terminologies. Our general idea is to partition the multiset $\mathcal{D}$ into several sub-multisets according to the way in which

---

**Algorithm 1:** DeepWalk

1. **for** $n = 1, 2, \ldots, N$ **do**
2.     Pick $w_1^n$ according to a probability distribution $P(w_1)$;
3.     Generate a vertex sequence $(w_1^n, \cdots, w_L^n)$ of length $L$ by a random walk on network $G$;
4.     **for** $j = 1, 2, \ldots, L - T$ **do**
5.         **for** $r = 1, \ldots, T$ **do**
6.             Add vertex-context pair $(w_j^n, w_{j+r}^n)$ to multiset $\mathcal{D}$;
7.             Add vertex-context pair $(w_{j+r}^n, w_j^n)$ to multiset $\mathcal{D}$;
8. Run SGNS on $\mathcal{D}$ with $b$ negative samples.

---

vertex and its context appear in a random walk sequence. More formally, for $r = 1, \cdots, T$, we define

$$\mathcal{D}_{\vec{r}} = \left\{ (w,c) : (w,c) \in \mathcal{D}, w = w_j^n, c = w_{j+r}^n \right\},$$

$$\mathcal{D}_{\overleftarrow{r}} = \left\{ (w,c) : (w,c) \in \mathcal{D}, w = w_{j+r}^n, c = w_j^n \right\}.$$

That is, $\mathcal{D}_{\vec{r}}/\mathcal{D}_{\overleftarrow{r}}$ is the sub-multiset of $\mathcal{D}$ such that the context $c$ is $r$ steps after/before the vertex $w$ in random walks. Moreover, we use $\#(w,c)_{\vec{r}}$ and $\#(w,c)_{\overleftarrow{r}}$ to denote the number of times vertex-context pair $(w,c)$ appears in $\mathcal{D}_{\vec{r}}$ and $\mathcal{D}_{\overleftarrow{r}}$, respectively. The following three theorems characterize DeepWalk step by step.

**Theorem 2.1.** Denote $P = D^{-1}A$, when $L \to \infty$, we have

$$\frac{\#(w,c)_{\vec{r}}}{|\mathcal{D}_{\vec{r}}|} \xrightarrow{p} \frac{d_w}{\operatorname{vol}(G)} (P^r)_{w,c} \text{ and } \frac{\#(w,c)_{\overleftarrow{r}}}{|\mathcal{D}_{\overleftarrow{r}}|} \xrightarrow{p} \frac{d_c}{\operatorname{vol}(G)} (P^r)_{c,w}.$$

**Proof.** See Appendix. □

**Remark 1.** *What if we start random walks with other distributions (e.g., the uniform distribution in the original DeepWalk work [31])? It turns out that, for a connected, undirected, and non-bipartite graph, $P(w_j = w, w_{j+r} = c) \to \frac{d_w}{\operatorname{vol}(G)} (P^r)_{w,c}$ as $j \to \infty$. So when the length of random walks $L \to \infty$, Thm. 2.1 still holds.*

**Theorem 2.2.** When $L \to \infty$, we have

$$\frac{\#(w,c)}{|\mathcal{D}|} \xrightarrow{p} \frac{1}{2T} \sum_{r=1}^{T} \left( \frac{d_w}{\operatorname{vol}(G)} (P^r)_{w,c} + \frac{d_c}{\operatorname{vol}(G)} (P^r)_{c,w} \right).$$

**Proof.** Note the fact that $\frac{|\mathcal{D}_{\vec{r}}|}{|\mathcal{D}|} = \frac{|\mathcal{D}_{\overleftarrow{r}}|}{|\mathcal{D}|} = \frac{1}{2T}$. By using Thm. 2.1 and the continuous mapping theorem, we get

$$\frac{\#(w,c)}{|\mathcal{D}|} = \frac{\sum_{r=1}^{T} \left( \#(w,c)_{\vec{r}} + \#(w,c)_{\overleftarrow{r}} \right)}{\sum_{r=1}^{T} \left( |\mathcal{D}_{\vec{r}}| + |\mathcal{D}_{\overleftarrow{r}}| \right)} = \frac{1}{2T} \sum_{r=1}^{T} \left( \frac{\#(w,c)_{\vec{r}}}{|\mathcal{D}_{\vec{r}}|} + \frac{\#(w,c)_{\overleftarrow{r}}}{|\mathcal{D}_{\overleftarrow{r}}|} \right)$$

$$\xrightarrow{p} \frac{1}{2T} \sum_{r=1}^{T} \left( \frac{d_w}{\operatorname{vol}(G)} (P^r)_{w,c} + \frac{d_c}{\operatorname{vol}(G)} (P^r)_{c,w} \right).$$

Further, marginalizing $w$ and $c$ respectively reveals the fact that $\frac{\#(w)}{|\mathcal{D}|} \xrightarrow{p} \frac{d_w}{\operatorname{vol}(G)}$ and $\frac{\#(c)}{|\mathcal{D}|} \xrightarrow{p} \frac{d_c}{\operatorname{vol}(G)}$, as $L \to \infty$. □

**Theorem 2.3.** *For DeepWalk, when $L \to \infty$,*

$$\frac{\#(w,c) |\mathcal{D}|}{\#(w) \cdot \#(c)} \xrightarrow{p} \frac{\operatorname{vol}(G)}{2T} \left( \frac{1}{d_c} \sum_{r=1}^{T} (P^r)_{w,c} + \frac{1}{d_w} \sum_{r=1}^{T} (P^r)_{c,w} \right).$$

*In matrix form, DeepWalk is equivalent to factorize*

$$\log\left(\frac{\text{vol}(G)}{T}\left(\sum_{r=1}^{T} P^r\right) D^{-1}\right) - \log(b). \quad (7)$$

PROOF. Using the results in Thm. 2.2 and the continuous mapping theorem, we get

$$\frac{\#(w,c)|\mathcal{D}|}{\#(w)\cdot\#(c)} = \frac{\frac{\#(w,c)}{|\mathcal{D}|}}{\frac{\#(w)}{|\mathcal{D}|}\cdot\frac{\#(c)}{|\mathcal{D}|}} \xrightarrow{p} \frac{\frac{1}{2T}\sum_{r=1}^{T}\left(\frac{d_w}{\text{vol}(G)}(P^r)_{w,c} + \frac{d_c}{\text{vol}(G)}(P^r)_{c,w}\right)}{\frac{d_w}{\text{vol}(G)}\cdot\frac{d_c}{\text{vol}(G)}}$$

$$= \frac{\text{vol}(G)}{2T}\left(\frac{1}{d_c}\sum_{r=1}^{T}(P^r)_{w,c} + \frac{1}{d_w}\sum_{r=1}^{T}(P^r)_{c,w}\right).$$

Write it in matrix form:

$$\frac{\text{vol}(G)}{2T}\left(\sum_{r=1}^{T} P^r D^{-1} + \sum_{r=1}^{T} D^{-1}(P^r)^\top\right)$$

$$= \frac{\text{vol}(G)}{2T}\left(\sum_{r=1}^{T} \underbrace{D^{-1}A \times \cdots \times D^{-1}A}_{r \text{ terms}} D^{-1} + \sum_{r=1}^{T} D^{-1}\underbrace{AD^{-1}\times\cdots\times AD^{-1}}_{r\text{ terms}}\right)$$

$$= \frac{\text{vol}(G)}{T}\sum_{r=1}^{T}\underbrace{D^{-1}A\times\cdots\times D^{-1}A}_{r\text{ terms}} D^{-1} = \text{vol}(G)\left(\frac{1}{T}\sum_{r=1}^{T}P^r\right)D^{-1}.$$

□

COROLLARY 2.4. *Comparing Eq. 4 and Eq. 7, we can easily observe that LINE (2nd) is a special case of DeepWalk when $T = 1$.*

## 2.3 node2vec

node2vec [16] is a recently proposed network embedding method, which is briefly summarized in Algorithm 2. First, it defines an unnormalized transition probability tensor $\underline{T}$ with parameters $p$ and $q$, and then normalizes it to be the transition probability of a 2nd-order random walk (Line 1). Formally,

$$\underline{T}_{u,v,w} = \begin{cases} \frac{1}{p} & (u,v)\in E, (v,w)\in E, u=w; \\ 1 & (u,v)\in E, (v,w)\in E, u\neq w, (w,u)\in E; \\ \frac{1}{q} & (u,v)\in E, (v,w)\in E, u\neq w, (w,u)\notin E; \\ 0 & \text{otherwise}. \end{cases}$$

$$\underline{P}_{u,v,w} = \text{Prob}\left(w_{j+1} = u|w_j = v, w_{j-1} = w\right) = \frac{\underline{T}_{u,v,w}}{\sum_u \underline{T}_{u,v,w}}.$$

Second, node2vec performs the 2nd-order random walks to generate a multiset $\mathcal{D}$ (Line 2-8) and then trains an SGNS model (Line 9) on it. To facilitate the analysis, we instead record triplets to form the multiset $\mathcal{D}$ for node2vec rather than vertex-context pairs. Take a vertex $w = w_j^n$ and its context $c = w_{j+r}^n$ as an example, we denote $u = w_{j-1}^n$ and add a triplet $(w,c,u)$ into $\mathcal{D}$ (Line 7). Similar to our analysis of DeepWalk, we partition the multiset $\mathcal{D}$ according to the way in which the vertex and its context appear in a random walk sequence. More formally, for $r = 1, \cdots, T$, we define

$$\mathcal{D}_{\vec{r}} = \left\{(w,c,u) : (w,c,u)\in \mathcal{D}, w_j^n = w, w_{j+r}^n = c, w_{j-1}^n = u\right\},$$

$$\mathcal{D}_{\overleftarrow{r}} = \left\{(w,c,u) : (w,c,u)\in \mathcal{D}, w_{j+r}^n = w, w_j^n = c, w_{j-1}^n = u\right\}.$$

In addition, for a triplet $(w,c,u)$, we use $\#(w,c,u)_{\vec{r}}$ and $\#(w,c,u)_{\overleftarrow{r}}$ to denote the number of times it appears in $\mathcal{D}_{\vec{r}}$ and $\mathcal{D}_{\overleftarrow{r}}$, respectively.

**Algorithm 2:** node2vec

1. Construct transition probability tensor $\underline{P}$;
2. **for** $n = 1, 2, \ldots, N$ **do**
3.    Pick $w_1^n, w_2^n$ according to a distribution $P(w_1, w_2)$;
4.    Generate a vertex sequence $(w_1^n, \cdots, w_L^n)$ of length $L$ by the 2nd-order random walk on network $G$;
5.    **for** $j = 2, 3, \ldots, L - T$ **do**
6.       **for** $r = 1, \ldots, T$ **do**
7.          Add triplet $(w_j^n, w_{j+r}^n, w_{j-1}^n)$ to multiset $\mathcal{D}$;
8.          Add triplet $(w_{j+r}^n, w_j^n, w_{j-1}^n)$ to multiset $\mathcal{D}$;
9. Run SGNS on multiset $\mathcal{D}' = \{(w,c) : (w,c,u)\in\mathcal{D}\}$;

In this analysis, we assume the first two vertices of node2vec's 2nd-order random walk are sampled from its stationary distribution $X$. The stationary distribution $X$ of the 2nd-order random walk satisfies $\sum_w \underline{P}_{u,v,w} X_{v,w} = X_{u,v}$, and the existence of such $X$ is guaranteed by the Perron-Frobenius theorem [4]. Additionally, the higher-order transition probability tensor is defined to be $(\underline{P}^r)_{u,v,w} = \text{Prob}(w_{j+r} = u|w_j = v, w_{j-1} = w)$.

**Main Results** Limited by space, we list the main results of node2vec without proofs. The idea is similar to the analysis of DeepWalk.

- $\frac{\#(w,c,u)_{\vec{r}}}{|\mathcal{D}_{\vec{r}}|} \xrightarrow{p} X_{w,u}(\underline{P}^r)_{c,w,u}$ and $\frac{\#(w,c,u)_{\overleftarrow{r}}}{|\mathcal{D}_{\overleftarrow{r}}|} \xrightarrow{p} X_{c,u}(\underline{P}^r)_{w,c,u}$.
- $\frac{\#(w,c)_{\vec{r}}}{|\mathcal{D}_{\vec{r}}|} = \frac{\sum_u \#(w,c,u)_{\vec{r}}}{|\mathcal{D}_{\vec{r}}|} \xrightarrow{p} \sum_u X_{w,u}(\underline{P}^r)_{c,w,u}$.
- $\frac{\#(w,c)_{\overleftarrow{r}}}{|\mathcal{D}_{\overleftarrow{r}}|} = \frac{\sum_u \#(w,c,u)_{\overleftarrow{r}}}{|\mathcal{D}_{\overleftarrow{r}}|} \xrightarrow{p} \sum_u X_{c,u}(\underline{P}^r)_{w,c,u}$.
- $\frac{\#(w,c)}{|\mathcal{D}|} \xrightarrow{p} \frac{1}{2T}\sum_{r=1}^{T}\left(\sum_u X_{w,u}(\underline{P}^r)_{c,w,u} + \sum_u X_{c,u}(\underline{P}^r)_{w,c,u}\right)$.
- $\frac{\#(w)}{|\mathcal{D}|} \xrightarrow{p} \sum_u X_{w,u}$ and $\frac{\#(c)}{|\mathcal{D}|} \xrightarrow{p} \sum_u X_{c,u}$.

Combining all together, we conclude that, for node2vec, Eq. 6 has the form

$$\frac{\#(w,c)|\mathcal{D}|}{\#(w)\cdot\#(c)} \xrightarrow{p} \frac{\frac{1}{2T}\sum_{r=1}^{T}\left(\sum_u X_{w,u}\underline{P}^r_{c,w,u} + \sum_u X_{c,u}\underline{P}^r_{w,c,u}\right)}{\left(\sum_u X_{w,u}\right)\left(\sum_u X_{c,u}\right)}. \quad (8)$$

Though the closed form for node2vec has been achieved, we leave the formulation of its matrix form for future research.

Note that both the computation and storage of the transition probability tensor $\underline{P}^r$ and its corresponding stationary distribution $X$ are very expensive, making the modeling of the full 2nd-order dynamics difficult. However, we have noticed some recent progresses [3, 4, 15] that try to understand or approximate the 2nd-order random walk by assuming a rank-one factorization $X_{u,v} = x_u x_v$ for its stationary distribution $X$. Due to the page limitation, in the rest of this paper, we mainly focus on the matrix factorization framework depending on the 1st-order random walk (DeepWalk).

## 3 NetMF: NETWORK EMBEDDING AS MATRIX FACTORIZATION

Based on the analysis in Section 2, we unify LINE, PTE, DeepWalk, and node2vec in the framework of matrix factorization, where the factorized matrices have closed forms as showed in Eq. 4, Eq. 5, Eq. 7, and Eq. 8, respectively. In this section, we study the DeepWalk matrix (Eq. 7) because it is more general than the LINE matrix and computationally more efficient than the node2vec matrix. We first discuss the connection between the DeepWalk matrix and graph Laplacian in Section 3.1. Then in Section 3.2, we present a matrix factorization based framework—NetMF—for network embedding.

## 3.1 Connection between DeepWalk Matrix and Normalized Graph Laplacian

In this section, we show that the DeepWalk matrix has a close relationship with the normalized graph Laplacian. To facilitate our analysis, we introduce the following four theorems.

THEOREM 3.1. ([11]) For normalized graph Laplacian $\mathcal{L} = I - D^{-1/2}AD^{-1/2} \in \mathbb{R}^{n \times n}$, all its eigenvalues are real numbers and lie in $[0, 2]$, with $\lambda_{\min}(\mathcal{L}) = 0$. For a connected graph with $n > 1$, $\lambda_{\max}(\mathcal{L}) \geq n/(n-1)$.

THEOREM 3.2. ([41]) Singular values of a real symmetric matrix are the absolute values of its eigenvalues.

THEOREM 3.3. ([19]) Let $B, C$ be two $n \times n$ symmetric matrices. Then for the decreasingly ordered singular values $\sigma$ of $B, C$ and $BC$, $\sigma_{i+j-1}(BC) \leq \sigma_i(B) \times \sigma_j(C)$ holds for any $1 \leq i, j \leq n$ and $i+j \leq n+1$.

THEOREM 3.4. ([41]) For a real symmetric matrix $A$, its Rayleigh Quotient $R(A, x) = (x^\top A x)/(x^\top x)$ satisfies $\lambda_{\min}(A) = \min_{x \neq 0} R(A, x)$ and $\lambda_{\max}(A) = \max_{x \neq 0} R(A, x)$.

By ignoring the element-wise matrix logarithm and the constant term in Eq. 7, we focus on studying the matrix $\left(\frac{1}{T}\sum_{r=1}^{T} P^r\right) D^{-1}$. By Thm. 3.1, $D^{-1/2}AD^{-1/2} = I - \mathcal{L}$ has eigen-decomposition $U\Lambda U^\top$ such that $U$ orthonormal and $\Lambda = \text{diag}(\lambda_1, \cdots, \lambda_n)$, where $1 = \lambda_1 \geq \lambda_2 \geq \cdots \geq \lambda_n \geq -1$ and $\lambda_n < 0$. Based on this eigen-decomposition, $\left(\frac{1}{T}\sum_{r=1}^{T} P^r\right) D^{-1}$ can be decomposed to be the product of there symmetric matrices:

$$\left(\frac{1}{T}\sum_{r=1}^{T} P^r\right) D^{-1} = \left(D^{-1/2}\right)\left(U\left(\frac{1}{T}\sum_{r=1}^{T} \Lambda^r\right)U^\top\right)\left(D^{-1/2}\right). \quad (9)$$

The goal here is to characterize the spectrum of $\left(\frac{1}{T}\sum_{r=1}^{T} P^r\right) D^{-1}$. To achieve this, we first analyze the second matrix at RHS of Eq. 9, and then extend our analysis to the targeted matrix at LHS.

**Spectrum of** $U\left(\frac{1}{T}\sum_{r=1}^{T} \Lambda^r\right)U^\top$ The matrix $U\left(\frac{1}{T}\sum_{r=1}^{T} \Lambda^r\right)U^\top$ has eigenvalues $\frac{1}{T}\sum_{r=1}^{T} \lambda_i^r, i = 1, \cdots, n$, which can be treated as the output of a transformation applied on $D^{-1/2}AD^{-1/2}$'s eigenvalue $\lambda_i$, i.e., a kind of filter! The effect of this transformation (filter) $f(x) = \frac{1}{T}\sum_{r=1}^{T} x^r$ is plotted in Figure 1(a), from which we observe the following two properties of this filter. Firstly, it prefers positive large eigenvalues; Secondly, the preference becomes stronger as the window size $T$ increases. In other words, as $T$ grows, this filter tries to approximate a low-rank positive semi-definite matrix by keeping large positive eigenvalues.

**Spectrum of** $\left(\frac{1}{T}\sum_{r=1}^{T} P^r\right) D^{-1}$ Guided by Thm. 3.2, the decreasingly ordered singular values of the matrix $U\left(\frac{1}{T}\sum_{r=1}^{T} \Lambda^r\right)U^\top$ can be constructed by sorting the absolute value of its eigenvalues in

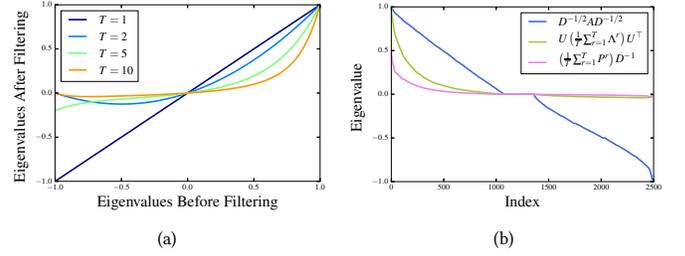

Figure 1: DeepWalk Matrix as Filtering: (a) Function $f(x) = \frac{1}{T}\sum_{r=1}^{T} x^r$ with dom $f = [-1, 1]$, where $T \in \{1, 2, 5, 10\}$; (b) Eigenvalues of $D^{-1/2}AD^{-1/2}$, $U\left(\frac{1}{T}\sum_{r=1}^{T} \Lambda^r\right)U^\top$, and $\left(\frac{1}{T}\sum_{r=1}^{T} P^r\right)D^{-1}$ for Cora network ($T = 10$).

the non-increasing order such that

$$\left|\frac{1}{T}\sum_{r=1}^{T}\lambda_{p_1}^r\right| \geq \left|\frac{1}{T}\sum_{r=1}^{T}\lambda_{p_2}^r\right| \geq \cdots \geq \left|\frac{1}{T}\sum_{r=1}^{T}\lambda_{p_n}^r\right|,$$

where $\{p_1, p_2, \cdots, p_n\}$ is a permutation of $\{1, 2, \cdots, n\}$. Similarly, since every $d_i$ is positive, the decreasingly ordered singular values of the matrix $D^{-1/2}$ can be constructed by sorting $1/\sqrt{d_i}$ in the non-increasing order such that $1/\sqrt{d_{q_1}} \geq 1/\sqrt{d_{q_2}} \geq \cdots \geq 1/\sqrt{d_{q_n}}$ where $\{q_1, q_2, \cdots, q_n\}$ is a permutation of $\{1, 2, \cdots, n\}$. In particular, $d_{q_1} = d_{\min}$ is the smallest vertex degree. By applying Thm. 3.3 twice, we can see that the $s$-th singular value satisfies

$$\sigma_s\left(\left(\frac{1}{T}\sum_{r=1}^{T} P^r\right)D^{-1}\right) \leq \sigma_1\left(D^{-\frac{1}{2}}\right)\sigma_s\left(U\left(\frac{1}{T}\sum_{r=1}^{T}\Lambda^r\right)U^\top\right)\sigma_1\left(D^{-\frac{1}{2}}\right)$$

$$= \frac{1}{\sqrt{d_{q_1}}}\left|\frac{1}{T}\sum_{r=1}^{T}\lambda_{p_s}^r\right|\frac{1}{\sqrt{d_{q_1}}} = \frac{1}{d_{\min}}\left|\frac{1}{T}\sum_{r=1}^{T}\lambda_{p_s}^r\right|, \quad (10)$$

which reveals that the magnitude of $\left(\frac{1}{T}\sum_{r=1}^{T} P^r\right)D^{-1}$'s eigenvalues is always bounded by the magnitude of $U\left(\frac{1}{T}\sum_{r=1}^{T}\Lambda^r\right)U^\top$'s eigenvalues. In addition to the magnitude of eigenvalues, we also want to bound its smallest eigenvalue. Observe that the Rayleigh Quotient of $\left(\frac{1}{T}\sum_{r=1}^{T} P^r\right)D^{-1}$ has a lower bound as follows

$$R\left(\left(\frac{1}{T}\sum_{r=1}^{T} P^r\right)D^{-1}, x\right) = R\left(U\left(\frac{1}{T}\sum_{r=1}^{T}\Lambda^r\right)U^\top, D^{-1/2}x\right) R\left(D^{-1}, x\right)$$

$$\geq \lambda_{\min}\left(U\left(\frac{1}{T}\sum_{r=1}^{T}\Lambda^r\right)U^\top\right)\lambda_{\max}\left(D^{-1}\right) = \frac{1}{d_{\min}}\lambda_{\min}\left(U\left(\frac{1}{T}\sum_{r=1}^{T}\Lambda^r\right)U^\top\right).$$

By applying Thm. 3.4, we can bound the smallest eigenvalue of $\left(\frac{1}{T}\sum_{r=1}^{T} P^r\right)D^{-1}$ by the smallest eigenvalue of $U\left(\frac{1}{T}\sum_{r=1}^{T}\Lambda^r\right)U^\top$:

$$\lambda_{\min}\left(\left(\frac{1}{T}\sum_{r=1}^{T} P^r\right)D^{-1}\right) \geq \frac{1}{d_{\min}}\lambda_{\min}\left(U\left(\frac{1}{T}\sum_{r=1}^{T}\Lambda^r\right)U^\top\right).$$

**Illustrative Example: Cora** In order to illustrate the filtering effect we discuss above, we analyze a small citation network Cora [27]. we make the citation links undirected and choose its largest connected component. In Figure 1(b), we plot the decreasingly ordered eigenvalues of matrices $D^{-1/2}AD^{-1/2}$, $U\left(\frac{1}{T}\sum_{r=1}^{T}\Lambda^r\right)U^\top$,

**Algorithm 3:** NetMF for a Small Window Size $T$

1 Compute $P^1, \cdots, P^T$;
2 Compute $M = \frac{\text{vol}(G)}{bT} \left( \sum_{r=1}^{T} P^r \right) D^{-1}$;
3 Compute $M' = \max(M, 1)$;
4 Rank-$d$ approximation by SVD: $\log M' = U_d \Sigma_d V_d^\top$;
5 **return** $U_d \sqrt{\Sigma_d}$ as network embedding.

---

**Algorithm 4:** NetMF for a Large Window Size $T$

1 Eigen-decomposition $D^{-1/2} A D^{-1/2} \approx U_h \Lambda_h U_h^\top$;
2 Approximate $M$ with
   $\hat{M} = \frac{\text{vol}(G)}{b} D^{-1/2} U_h \left( \frac{1}{T} \sum_{r=1}^{T} \Lambda_h^r \right) U_h^\top D^{-1/2}$;
3 Compute $\hat{M}' = \max(\hat{M}, 1)$;
4 Rank-$d$ approximation by SVD: $\log \hat{M}' = U_d \Sigma_d V_d^\top$;
5 **return** $U_d \sqrt{\Sigma_d}$ as network embedding.

---

and $\left( \frac{1}{T} \sum_{r=1}^{T} P^r \right) D^{-1}$, respectively, with $T = 10$. For $D^{-1/2} A D^{-1/2}$, the largest eigenvalue $\lambda_1 = 1$, and the smallest eigenvalue $\lambda_n = -0.971$. For $U \left( \frac{1}{T} \sum_{r=1}^{T} \Lambda^r \right) U^\top$, we observe that all its negative eigenvalues and small positive eigenvalues are "filtered out" in spectrum. Finally, for the matrix $\left( \frac{1}{T} \sum_{r=1}^{T} P^r \right) D^{-1}$, we observe that both the magnitude of its eigenvalues and its smallest eigenvalue are bounded by those of $U \left( \frac{1}{T} \sum_{r=1}^{T} \Lambda^r \right) U^\top$.

## 3.2 NetMF

Built upon the theoretical analysis above, we propose a matrix factorization framework NetMF for empirical understanding of and improving on DeepWalk and LINE. For simplicity, we denote $M = \frac{\text{vol}(G)}{bT} \left( \sum_{r=1}^{T} P^r \right) D^{-1}$, and refer to $\log M$ as the DeepWalk matrix.

**NetMF for a Small Window Size $T$** NetMF for a small $T$ is quite intuitive. The basic idea is to directly compute and factorize the DeepWalk matrix. The detail is listed in Algorithm 3. In the first step (Line 1-2), we compute the matrix power from $P^1$ to $P^T$ and then get $M$. However, the factorization of $\log M$ presents computational challenges due to the element-wise matrix logarithm. The matrix is not only ill-defined (since $\log 0 = -\infty$), but also dense. Inspired by the *Shifted* PPMI approach [24], we define $M'$ such that $M'_{i,j} = \max(M_{i,j}, 1)$ (Line 3). In this way, $\log M'$ is a sparse and consistent version of $\log M$. Finally, we factorize $\log M'$ by using Singular Value Decomposition (SVD) and construct network embedding by using its top-$d$ singular values/vectors (Line 4-5).

**NetMF for a Large Window Size $T$** The direct computation of the matrix $M$ presents computing challenges for a large window size $T$, mainly due to its high time complexity. Hereby we propose an approximation algorithm as listed in Algorithm 4. The general idea comes from our analysis in section 3.1, wherein we reveal $M$'s close relationship with the normalized graph Laplacian and show its low-rank nature theoretically. In our algorithm, we first approximate $D^{-1/2} A D^{-1/2}$ with its top-$h$ eigenpairs $U_h \Lambda_h U_h^\top$ (Line 1). Since only the top-$h$ eigenpairs are required and the involved matrix is sparse, we can use Arnoldi method [22] to achieve significant time reduction. In step two (Line 2), we approximate $M$ with $\hat{M} = \frac{\text{vol}(G)}{b} D^{-1/2} U_h \left( \frac{1}{T} \sum_{r=1}^{T} \Lambda_h^r \right) U_h^\top D^{-1/2}$. The final step is the same as that in Algorithm 3, in which we form $\hat{M}' = \max(\hat{M}, 1)$ (Line 3) and then perform SVD on $\log \hat{M}'$ to get network embedding (Line 4-5).

**Table 2: Statistics of Datasets.**

| Dataset | BlogCatalog | PPI | Wikipedia | Flickr |
|---|---|---|---|---|
| $|V|$ | 10,312 | 3,890 | 4,777 | 80,513 |
| $|E|$ | 333,983 | 76,584 | 184,812 | 5,899,882 |
| #Labels | 39 | 50 | 40 | 195 |

For NetMF with large window sizes, we develop the following error bound theorem for the approximation of $M$ and the approximation of $\log M'$.

THEOREM 3.5. *Let $\|\cdot\|_F$ be the matrix Frobenius norm. Then*

$$\left\| M - \hat{M} \right\|_F \leq \frac{\text{vol}(G)}{b d_{\min}} \sqrt{\sum_{j=k+1}^{n} \left| \frac{1}{T} \sum_{r=1}^{T} \lambda_j^r \right|^2};$$

$$\left\| \log M' - \log \hat{M}' \right\|_F \leq \left\| M' - \hat{M}' \right\|_F \leq \left\| M - \hat{M} \right\|_F.$$

PROOF. See Appendix. □

Thm. 3.5 reveals that the error for approximating $\log M'$ is bounded by the error bound for the approximation of $M$. Nevertheless, the major drawback of NetMF lies in this element-wise matrix logarithm. Since good tools are currently not available to analyze this operator, we have to compute it explicitly even after we have already achieved a good low-rank approximation of $M$.

## 4 EXPERIMENTS

In this section, we evaluate the proposed NetMF method on the multi-label vertex classification task, which has also been used in the works of DeepWalk, LINE, and node2vec.

**Datasets** We employ four widely-used datasets for this task. The statistics of these datasets are listed in Table 2.

**BlogCatalog** [38] is a network of social relationships of online bloggers. The vertex labels represent interests of the bloggers.

**Protein-Protein Interactions (PPI)** [35] is a subgraph of the PPI network for Homo Sapiens. The labels are obtained from the hallmark gene sets and represent biological states.

**Wikipedia**[3] is a co-occurrence network of words appearing in the first million bytes of the Wikipedia dump. The labels are the Part-of-Speech (POS) tags inferred by Stanford POS-Tagger [40].

**Flickr** [38] is the user contact network in Flickr. The labels represent the interest groups of the users.

**Baseline Methods** We compare our methods NetMF ($T = 1$) and NetMF ($T = 10$) with LINE (2nd) [37] and DeepWalk [31], which we have introduced in previous sections. For NetMF ($T = 10$), we choose $h = 16384$ for Flickr, and $h = 256$ for BlogCatelog, PPI, and

---

[3]http://mattmahoney.net/dc/text.html

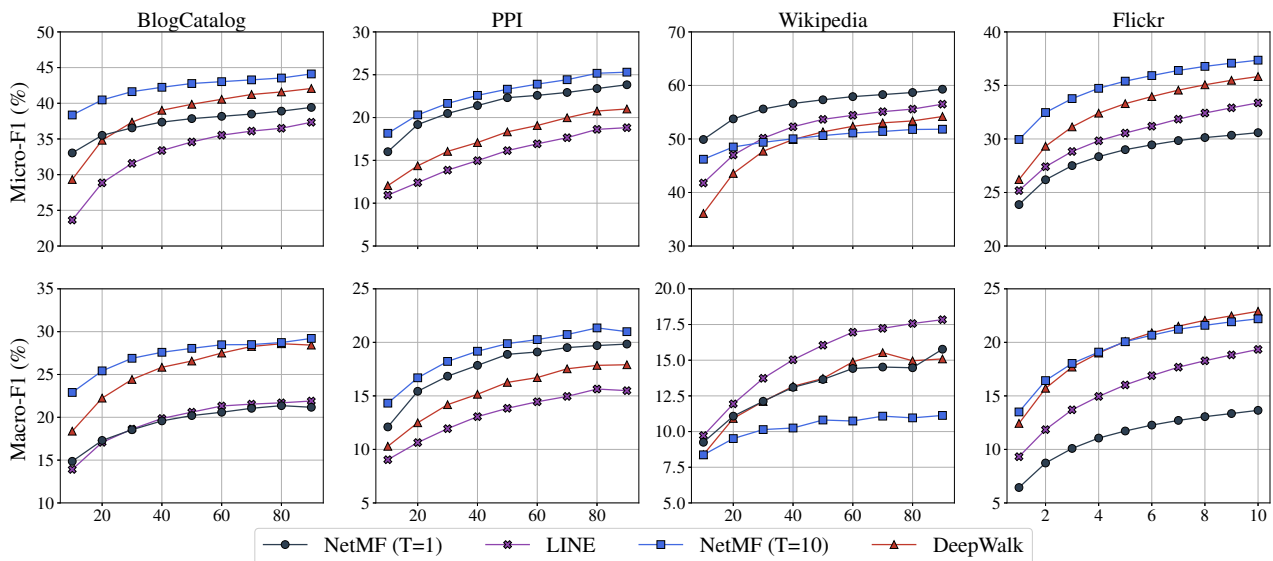

Figure 2: Predictive performance on varying the ratio of training data. The $x$-axis represents the ratio of labeled data (%), and the $y$-axis in the top and bottom rows denote the Micro-F1 and Macro-F1 scores respectively.

Wikipedia. For DeepWalk, we present its results with the authors' preferred parameters—window size 10, walk length 40, and the number of walks starting from each vertex to be 80. Finally, we set embedding dimension to be 128 for all methods.

**Prediction Setting** Following the same experimental procedure in DeepWalk [31], we randomly sample a portion of labeled vertices for training and use the rest for testing. For BlogCatalog, PPI, and Wikipedia datasets, the training ratio is varied from 10% to 90%. For Flickr, the training ratio is varied from 1% to 10%. We use the one-vs-rest logistic regression model implemented by LIBLINEAR [14] for the multi-label classification task. In the test phase, the one-vs-rest model yields a ranking of labels rather than an exact label assignment. To avoid the thresholding effect [39], we assume that the number of labels for test data is given [31, 39]. We repeat the prediction procedure 10 times and evaluate the performance in terms of average Micro-F1 and average Macro-F1 [42].

**Experimental Results** Figure 2 summarizes the prediction performance of all methods on the four datasets and Table 3 lists the quantitative and relative gaps between our methods and baselines. In specific, we show NetMF ($T = 1$)'s relative performance gain over LINE (2nd) and NetMF ($T = 10$)'s relative improvements over DeepWalk, respectively, as each pair of them share the same window size $T$. We have the following key observations and insights:

(1) In BlogCatalog, PPI, and Flickr, the proposed NetMF method ($T = 10$) achieves significantly better predictive performance over baseline approaches as measured by both Micro-F1 and Macro-F1, demonstrating the effectiveness of the theoretical foundation we lay out for network embedding.

(2) In Wikipedia, NetMF ($T = 1$) shows better performance than other methods in terms of Micro-F1, while LINE outperforms other methods regarding Macro-F1. This observation implies that short-term dependence is enough to model Wikipedia's network structure.

This is because the used Wikipedia network is a dense word co-occurrence network with the average degree = 77.11, in which an edge between a pair of words are connected if they co-occur in a two-length window in the Wikipedia corpus.

(3) As shown in Table 3, the proposed NetMF method ($T = 10$ and $T = 1$) outperforms DeepWalk and LINE by large margins in most cases when sparsely labeled vertices are provided. Take the PPI dataset with 10% training data as an example, NetMF ($T = 1$) achieves relatively 46.34% and 33.85% gains over LINE (2nd) regarding Micro-F1 and Macro-F1 scores, respectively; More impressively, NetMF ($T = 10$) outperforms DeepWalk by 50.71% and 39.16% relatively as measured by two metrics.

(4) DeepWalk tries to approximate the exact vertex-context joint distribution with an empirical distribution through random walk sampling. Although the convergence is guaranteed by the law of large numbers, there still exist gaps between the exact and estimated distributions due to the large size of real-world networks and the relatively limited scale of random walks in practice (e.g., #walks and the walk length), negatively affecting DeepWalk's performance.

## 5 RELATED WORK

The story of network embedding stems from Spectral Clustering [5, 45], a data clustering technique which selects eigenvalues/eigenvectors of a data affinity matrix to obtain representations that can be clustered or embedded in a low-dimensional space. Spectral Clustering has been widely used in fields such as community detection [23] and image segmentation [33]. In recent years, there is an increasing interest in network embedding. Following a few pioneer works such as SocDim [38] and DeepWalk [31], a growing number of literature has tried to address the problem from various of perspectives, such as heterogeneous network embedding [8, 12, 20, 36], semi-supervised network embedding [17, 21, 44, 48], network embedding with rich vertex attributes [43, 47, 49], network embedding with high order structure [6, 16], signed network embedding [10], direct network embedding [30], network embedding via deep neural network [7, 25, 46], etc.

Table 3: Micro/Macro-F1 Score(%) for Multilabel Classification on BlogCatalog, PPI, Wikipedia, and Flickr datasets. In Flickr, 1% of vertices are labeled for training [31], and in the other three datasets, 10% of vertices are labeled for training.

| Algorithm | BlogCatalog (10%) | | PPI (10%) | | Wikipeida (10%) | | Flickr (1%) | |
|---|---|---|---|---|---|---|---|---|
| | Micro-F1 | Macro-F1 | Micro-F1 | Macro-F1 | Micro-F1 | Macro-F1 | Micro-F1 | Macro-F1 |
| LINE (2nd) | 23.64 | 13.91 | 10.94 | 9.04 | 41.77 | **9.72** | 25.18 | 9.32 |
| NetMF ($T = 1$) | 33.04 | 14.86 | 16.01 | 12.10 | **49.90** | 9.25 | 23.87 | 6.44 |
| Relative Gain of NetMF ($T = 1$) | 39.76 | 6.83% | 46.34% | 33.85% | 19.46% | -4.84% | -5.20% | -30.90% |
| DeepWalk | 29.32 | 18.38 | 12.05 | 10.29 | 36.08 | 8.38 | 26.21 | 12.43 |
| NetMF ($T = 10$) | **38.36** | **22.90** | **18.16** | **14.32** | 46.21 | 8.38 | **29.95** | **13.50** |
| Relative Gain of NetMF ($T = 10$) | 30.83% | 24.59% | 50.71% | 39.16% | 28.08% | 0.00% | 14.27% | 8.93% |

Among the above research, a commonly used technique is to define the "context" for each vertex, and then to train a predictive model to perform context prediction. For example, DeepWalk [31], node2vec [16], and metapath2vec [12] define vertices' context by the 1st-, 2nd-order, and meta-path based random walks, respectively; The idea of leveraging the context information are largely motivated by the skip-gram model with negative sampling (SGNS) [29]. Recently, there has been effort in understanding this model. For example, Levy and Goldberg [24] prove that SGNS is actually conducting an implicit matrix factorization, which provides us with a tool to analyze the above network embedding models; Arora et al. [1] propose a generative model RAND-WALK to explain word embedding models; and Hashimoto et al. [18] frame word embedding as a metric learning problem. Built upon the work in [24], we theoretically analyze popular skip-gram based network embedding models and connect them with spectral graph theory.

## 6 CONCLUSION

In this work, we provide a theoretical analysis of four impactful network embedding methods—DeepWalk, LINE, PTE, and node2vec—that were recently proposed between the years 2014 and 2016. We show that all of the four methods are essentially performing implicit matrix factorizations and the closed forms of their matrices offer not only the relationships between those methods but also their intrinsic connections with graph Laplacian. We further propose NetMF—a general framework to explicitly factorize the closed-form matrices that DeepWalk and LINE aim to implicitly approximate and factorize. Our extensive experiments suggest that NetMF's direct factorization achieves consistent performance improvements over the implicit approximation models—DeepWalk and LINE.

In the future, we would like to further explore promising directions to deepen our understanding of network embedding. It would be necessary to investigate whether and how the development in random-walk polynomials [9] can support fast approximations of the closed-form matrices. The computation and approximation of the 2nd-order random walks employed by node2vec is another interesting topic to follow. Finally, it is exciting to study the nature of skip-gram based dynamic and heterogeneous network embedding.

**Acknowledgements.** We thank Hou Pong Chan for his comments. Jiezhong Qiu and Jie Tang are supported by NSFC 61561130160. Jian Li is supported in part by the National Basic Research Program of China Grant 2015CB358700, and NSFC 61772297 & 61632016.

## APPENDIX:

THEOREM 2.1. *Denote* $P = D^{-1}A$, *when* $L \to \infty$, *we have*

$$\frac{\#(w, c)_{\overrightarrow{}}}{|\mathcal{D}_{\overrightarrow{}}|} \xrightarrow{p} \frac{d_w}{\text{vol}(G)} (P^r)_{w,c} \text{ and } \frac{\#(w, c)_{\overleftarrow{}}}{|\mathcal{D}_{\overleftarrow{}}|} \xrightarrow{p} \frac{d_c}{\text{vol}(G)} (P^r)_{c,w}.$$

LEMMA 6.1. *(S.N. Bernstein Law of Large Numbers [34]) Let* $Y_1, Y_2 \cdots$ *be a sequence of random variables with finite expectation* $\mathbb{E}[Y_j]$ *and variance* $\text{Var}(Y_j) < K, j \geq 1$, *and covariances are s.t.* $\text{Cov}(Y_i, Y_j) \to 0$ *as* $|i - j| \to \infty$. *Then the law of large numbers (LLN) holds.*

PROOF. First consider the special case when $N = 1$, thus we only have one vertex sequence $w_1, \cdots, w_L$ generated by random walk as described in Algorithm 1. Consider one certain vertex-context pair $(w, c)$, let $Y_j$ $(j = 1, \cdots, L - T)$ be the indicator function for event that $w_j = w$ and $w_{j+r} = c$. We have the following two observations:

- The quantity $\#(w, c)_{\overrightarrow{}}/|\mathcal{D}_{\overrightarrow{}}|$ is the sample average of $Y_j$'s, i.e.,

$$\frac{\#(w, c)_{\overrightarrow{}}}{|\mathcal{D}_{\overrightarrow{}}|} = \frac{1}{L - T} \sum_{j=1}^{L-T} Y_j.$$

- Based on our assumptions about the graph and the random walk, $\mathbb{E}[Y_j] = \frac{d_w}{\text{vol}(G)} (P^r)_{w,c}$, and when $j > i + r$,

$$\mathbb{E}(Y_i Y_j) = \text{Prob}(w_i = w, w_{i+r} = c, w_j = w, w_{j+r} = c)$$

$$= \frac{d_w}{\text{vol}(G)} (P^r)_{w,c} \left(P^{j-i+r}\right)_{c,w} (P^r)_{w,c}.$$

In this way, we can evaluate the covariance $\text{Cov}(Y_i, Y_j)$ when $j > i + r$ and calculate its limit when $j - i \to \infty$ by using the fact that our random walk converges to its stationary distribution:

$$\text{Cov}(Y_i, Y_j) = \mathbb{E}(Y_i Y_j) - \mathbb{E}(Y_i)\mathbb{E}(Y_j)$$

$$= \frac{d_w}{\text{vol}(G)} (P^r)_{w,c} \underbrace{\left(\left(P^{j-i+r}\right)_{c,w} - \frac{d_w}{\text{vol}(G)}\right)}_{\text{goes to 0 as } j-i \to \infty} (P^r)_{w,c} \to 0.$$

Then we can apply Lemma 6.1 and conclude that the sample average converges in probability towards the expected value, i.e.,

$$\frac{\#(w, c)_{\overrightarrow{}}}{|\mathcal{D}_{\overrightarrow{}}|} = \frac{1}{L - T} \sum_{j=1}^{L-T} Y_j \xrightarrow{p} \frac{1}{L - T} \sum_{j=1}^{L-T} \mathbb{E}(Y_j) = \frac{d_w}{\text{vol}(G)} (P^r)_{w,c}.$$

Similarly, we also have $\frac{\#(w,c)_{\overleftarrow{}}}{|\mathcal{D}_{\overleftarrow{}}|} \xrightarrow{p} \frac{d_c}{\text{vol}(G)} (P^r)_{c,w}$.

For the general case when $N > 1$, we define $Y_j^n$ ($n = 1, \cdots, N$, $j = 1, \cdots, L - T$) to be the indicator function for event $w_j^n = w$ and $w_{j+r}^n = c$, and organize $Y_j^n$'s as $Y_1^1, Y_1^2, \cdots, Y_1^N, Y_2^1, Y_2^2, \cdots, Y_2^N, \cdots$. This r.v. sequence still satisfies the condition of S.N. Bernstein LLN, so the above conclusion still holds. □

Theorem 3.5. *Let $\|\cdot\|_F$ be the matrix Frobenius norm. Then*

$$\left\|M - \hat{M}\right\|_F \le \frac{\text{vol}(G)}{b d_{\min}} \sqrt{\sum_{j=k+1}^{n} \left|\frac{1}{T}\sum_{r=1}^{T} \lambda_j^r\right|^2};$$

$$\left\|\log M' - \log \hat{M}'\right\|_F \le \left\|M' - \hat{M}'\right\|_F \le \left\|M - \hat{M}\right\|_F.$$

Proof. The first inequality can be seen by applying the definition of Frobenius norm and Eq. 10.

For the second inequality, first to show $\left\|\log M' - \log \hat{M}'\right\|_F \le \left\|M' - \hat{M}'\right\|_F$. According to the definition of Frobenius norm, sufficient to show $\left|\log M'_{i,j} - \log \hat{M}'_{i,j}\right| \le \left|\hat{M}'_{i,j} - M'_{i,j}\right|$ for any $i, j$. Without loss of generality, assume $M'_{i,j} \le \hat{M}'_{i,j}$.

$$\left|\log M'_{i,j} - \log \hat{M}'_{i,j}\right| = \log \frac{\hat{M}'_{i,j}}{M'_{i,j}} = \log\left(1 + \frac{\hat{M}'_{i,j} - M'_{i,j}}{M'_{i,j}}\right)$$

$$\le \frac{\hat{M}'_{i,j} - M'_{i,j}}{M'_{i,j}} \le \hat{M}'_{i,j} - M'_{i,j} = \left|\hat{M}'_{i,j} - M'_{i,j}\right|,$$

where the first inequality is because $\log(1+x) \le x$ for $x \ge 0$, and the the second inequality is because $M'_{i,j} = \max(M_{i,j}, 1) \ge 1$. Next to show $\left\|M' - \hat{M}'\right\|_F \le \left\|M - \hat{M}\right\|_F$. Sufficient to show $\left|M'_{i,j} - \hat{M}'_{i,j}\right| \le \left|M_{i,j} - \hat{M}_{i,j}\right|$ for any $i, j$. Recall the definition of $M'$ and $\hat{M}'$, we get $\left|M'_{i,j} - \hat{M}'_{i,j}\right| = \left|\max(M_{i,j}, 1) - \max(\hat{M}_{i,j}, 1)\right| \le \left|M_{i,j} - \hat{M}_{i,j}\right|$. □